\documentclass[prX,aps,tightenlines,preprint,floats,epsfig,superscriptaddress,nofootinbib]{revtex4}
\usepackage{amsfonts}
\usepackage{mathrsfs}
\usepackage{amssymb}
\usepackage{color}
\usepackage{epsfig}
\usepackage{graphicx}
\usepackage{amsmath}
\usepackage{cancel}

\begin{document}
\begin{titlepage}


\title{\Large MSSM in view of PAMELA and Fermi-LAT }

\author{Borut~Bajc}
\affiliation{J.\ Stefan Institute, 1000 Ljubljana, Slovenia}
\author{Tsedenbaljir~Enkhbat}
\affiliation{International Centre for Theoretical Physics, 34100
Trieste, Italy} \affiliation{J.\ Stefan Institute, 1000 Ljubljana,
Slovenia}
\author{Dilip~Kumar~Ghosh}
\affiliation{Department of Theoretical Physics and Centre for
Theoretical Sciences, Indian Association for the Cultivation of
Science, 2A \& 2B Raja S.C. Mullick Road, Kolkata 700 032, India}
 \author{Goran~Senjanovi\' c}
\affiliation{International Centre for Theoretical Physics, 34100
Trieste, Italy}
\author{Yue Zhang}
\affiliation{International Centre for Theoretical Physics, 34100
Trieste, Italy}


\begin{abstract}
We take the MSSM as a complete theory of low energy phenomena, including
neutrino masses and mixings. This immediately implies that the
gravitino is the only possible dark matter candidate. We study the
implications of the astrophysical experiments such as PAMELA and Fermi-LAT,
on this scenario. 
The theory can account for both the realistic neutrino masses and mixings,
 and the PAMELA data as long as the slepton masses lie in the $500-10^6\,$TeV range.
The squarks can be either light or heavy, depending on their contribution to radiative
neutrino masses.  
 On the other hand, the Fermi-LAT data imply heavy superpartners, all out of
LHC reach, simply on the grounds of the energy scale involved, for the gravitino must
weigh more than 2\,TeV. The perturbativity of the theory also implies
an upper bound on its mass, approximately $6-7\,$TeV. 
\end{abstract}


\maketitle
\thispagestyle{empty}
\end{titlepage}

\newpage
\section{Introduction}

The minimal supersymmetric standard model (MSSM)~\cite{Martin:1997ns} has become over the years the principal 
extension of the standard model. For this reason we study the
consequences of the possibility of MSSM being a complete theory of
present-day  phenomena. It has all the necessary ingredients to be
that: (i) it can naturally provide neutrino mass and mixings (unless
one artificially forbids many gauge invariant couplings by imposing
the so-called R-parity); (ii) if there is a light stop, it is
tailor-fit for electroweak baryogenesis~\cite{Carena:2008vj}, and it
also allows the Affleck-Dine baryogenesis through flat directions~\cite{Affleck:1984fy}; (iii) the same flat directions can provide a
natural source of inflation~\cite{Allahverdi:2006iq}; (iv) last but not least, it has a
number of neutral particles that can in principle be dark matter
candidates~\cite{Jungman:1995df}. This last issue is the subject of
the present undertaking. Our work is inspired by the recent
satellite experiments such as PAMELA~\cite{Adriani:2008zq}, ATIC and
Fermi-LAT~\cite{Abdo:2009zk} whose measurement on the high-energy
positron/electron excess has attracted a lot of attention.

\medskip

Here we define the MSSM as the supersymmetric extension of the minimal standard model,
without any new particles but the gravitino. The parameter space of the theory will be left open and subject to the experimental 
determination. In particular we make no assumption about the soft supersymmetry breaking terms, that is
superpartner masses and their mixings.
Taking the MSSM to be the complete low energy theory leads to a number of 
immediate consequences.  First, the only possible dark
matter candidate is precisely the gravitino. Second,
the gravitino decay through R-parity violation can account for the PAMELA data,
provided that the R-parity violation (RPV)~\cite{Barbier:2004ez} is hadrophobic and gravitino weights at least 300 GeV. 
The neutrino masses can also be explained
through radiative R-parity breaking corrections. Third, Fermi-LAT data can
be simultaneously explained through the gravitino decay as long as its mass is
greater than $2-3\,$TeV. This would unfortunately make all the
superpartners too heavy to be discovered at the LHC. 

\medskip

There is nothing original in our taking gravitino as dark matter, or even as a decaying 
dark matter~\cite{Takayama:2000uz}~\cite{Buchmuller:2009xv}~\cite{Bomark:2009zm}~\cite{Buchmuller:2009fm}.
In the case under study, this is forced on us, simply by taking MSSM as also a theory 
of neutrino masses and mixings.  Without this requirement, there would be 
no constraint on the MSSM parameters and the sfermions could be as light as one wishes,
as long as gravitino remains the LSP. The other studies tend to extend the MSSM,
 normally for the sake of baryogenesis and/or neutrino masses, which to us appears unnecessary. 

\medskip

Our main findings are the following.
\begin{itemize}
\item  Barring fine-tuned cancellations, in order to explain PAMELA results, sleptons have to
weigh between $500 - 10^6\,$TeV, which in turn implies no
observable lepton flavor violation in near future.
For a moderately light gravitino, $m_{3/2}\leq400-500\,$GeV, which corresponds to the parameter space
of the MSSM relevant for the LHC, the upper bound on sleptons masses
goes down to $10^4\,$TeV. 
 For Fermi-LAT, the slepton mass range becomes $10^4 - 10^6\,$TeV.
 
\item  Although R-parity breaking associated with the quark sector is subdominant
in gravitino decay compared to that in the lepton sector, it may still play an
important role in the neutrino mass. In this case, one would also end up with
heavy squarks, with similar lower limits on their masses as for the sleptons. 
This fits nicely with a split supersymmetry picture.   

 \item An interesting finding emerges if the the next-to-lightest superpartner (NLSP) is a wino/bino. 
 It decays through R-parity violation and its lifetime is rather long, about  $10^{-7}\,$sec, since the sleptons are heavy.
 At the Large Hadron Collider (LHC), once produced, it would still decay inside the detector~\cite{Ishiwata:2008tp}, producing multi-lepton finals states.
\end{itemize}

Before we turn to the detailed study, let us comment on the implications of our results for the two main
motivations for the low-energy supersymmetry.

\medskip

\paragraph* {The stabilization of gauge hierarchy.}
This requires light
supersymmetric partners. In particular the stop should weigh less than TeV,
while the other sfermion masses depend on the associated Yukawa couplings.
It is worth noting that, with a lower end mass needed to explain the PAMELA data,
even stau barely destabilizes the Higgs mass naturalness, for small values of $\tan \beta$.
  In the case of Fermi-LAT, naturalness is 
 gone completely. There would be no reason to worry about it, since the necessarily
 large gravitino mass, above a few TeV kills the MSSM as a theory verifiable at the LHC.

\paragraph*{The unification of gauge couplings.} This is a great success of the
MSSM for it was predicted~\cite{Dimopoulos:1981yj}~\cite{Ibanez:1981yh} \cite{Einhorn:1981sx}
 \cite{Marciano:1981un}
ten years before the LEP measurement
of the weak mixing angle. It also made prophetically a case for a large top Yukawa coupling~\cite{Marciano:1981un}. The unification works for arbitrarily large sfermion masses
 (as long as they are approximately equal within each generation), as in the case of
  split supersymmetry~\cite{ArkaniHamed:2004fb}. The heavy sleptons needed for PAMELA
  and Fermi-LAT thus bring some tension between naturalness and uniÞcation.
 

\section{MSSM implies decaying gravitino dark matter}

Besides the supersymmetric generalization of the standard model, the
most general gauge invariant MSSM contains the following
renormalizable interactions:
\begin{eqnarray}
W_{\cancel R} = \frac{1}{2} \lambda L L e^c + \lambda' Q L d^c +
\frac{1}{2} \lambda'' u^c d^c d^c + \mu' L H_u \ .
\end{eqnarray}
For simplicity we suppress the indices for families. These
interactions are often set {\it ad hoc} to zero, by assuming the
so-called R-parity.
Notice that supersymmetry by itself guarantees the stability of these
interactions set to zero. Notice also that zero is neither a special
point, nor is it physically preferred. We take the attitude here
that experiment should decide the values of these couplings, and if
MSSM is to be a complete theory, at least some of them must be
non-vanishing. Otherwise neutrinos will be massless. 
 
The third term in Eq.~(1) breaks baryon number
while the other terms break lepton number. 
The most stringent constraint comes from the proton decay~\cite{Smirnov:1996bg}: 
\begin{equation}
\lambda'\lambda''\lesssim10^{-27}
\left(\frac{m_{\widetilde d}}{300\rm GeV}\right)^2.
\end{equation} 
There is also the constraint from neutron-antineutron
oscillation~\cite{Dimopoulos:1987rk} 
\begin{equation}
\lambda''\lesssim
(10^{-7}-10^{-8})\left(\frac{m_{\widetilde d}}{100\rm
GeV}\right)^2\left(\frac{m_{\widetilde \chi^0}}{100\rm
GeV}\right)^{1/2}.
\end{equation} 
 Upon supersymmetry breaking, in general one
expects a non-vanishing sneutrino vacuum
expectation value (VEV), through a lepton number breaking soft term (corresponding to
the $\mu'$ term in Eq.~(1)). 

The neutrino masses can be generated either through $\lambda$
and/or $\lambda'$ at one-loop level or by sneutrino VEV
$\langle\widetilde \nu\rangle$ at tree level,
\begin{eqnarray}
m_\nu \simeq \displaystyle 
\displaystyle\frac{\lambda^2 (m_{\widetilde \ell}^2)_{LR}
m_\tau}{16\pi^2 m_{\widetilde \ell}^2}; \ \ \ \displaystyle\frac{3
\lambda'^2 (m_{\widetilde q}^2)_{LR}  m_b}{16\pi^2 m_{\widetilde
q}^2}; \ \ \ \frac{g^2 \langle\widetilde
\nu\rangle^2}{m_{\widetilde \chi^0}}\ ,
\end{eqnarray}
where  $m_{\widetilde \ell}$, $m_{\widetilde q}$ are the soft
masses for sleptons and squarks, respectively and  $(m_{\widetilde
f}^2)_{LR}$
are the mass-squared mixings
between left- and right-handed sfermions. We take here for simplicity left- and right-sleptons 
to be mass eigenstates with the same mass,  
$m_{\widetilde\ell}^2\equiv m_{\widetilde L}^2 \simeq m_{\widetilde e^c}^2$ and similarly for squarks
$m_{\widetilde q}^2\equiv m_{\widetilde Q}^2 \simeq m_{\widetilde d^c}^2$.
It is possible that one of these states is much heavier than the other, which is of 
less physical interest, so we postpone it for the subsection where we 
address the realistic situation in the general case.
Again the family indices are suppressed. The
reason that $\tau$ lepton and bottom quark masses are selected is
because they make the dominant contribution to neutrino mass,
barring accidental cancellations and fine-tunings. In general, it is
difficult to quantify individual contributions due to the
possibility of destructive/constructive interference between them.
It is instructive to explore the possibility that a particular term
dominates the neutrino mass.
\begin{itemize}
\item $\lambda$ dominates: here we have the following
\begin{eqnarray}\lambda\simeq 10^{-3}\left(\frac{m_{\widetilde{\ell}}}{1\mbox{TeV}}\right)
\left(\frac{m_\nu}{0.1\,{\rm eV}}\right)^{1/2}
\left(\frac{(m^2_{\widetilde{\ell}})_{\rm LR}}{(100\,\mbox{GeV})^2}\right)^{-1/2} \ .
\end{eqnarray} 
If this were the sole source for neutrino mass, it could be considered as appealing radiative seesaw mechanism
that produces light neutrinos for moderately small values of the ``Yukawa" coupling $\lambda$.
From here, one gets roughly a lower bound $\lambda\geq 10^{-4}$. As we will see later, PAMELA
forces it to be essentially larger.
\item $\lambda^\prime$ dominates: for the last option, we have a similar expression for the 
$\lambda^\prime$--coupling as $\lambda$ which differs only by a factor $\sim1/3$ due to 
the color and $m_b$  in Eq.~(2) for the same choice of sfermion mass parameters.
\item $\langle\widetilde
\nu\rangle$ dominates: in this case one get a seesaw mechanism for neutrino mass
where the usual right-handed neutrino gets traded for a gaugino, such as photino.
This implies a mixing between neutrino and gaugino, by analogy of the left-handed 
and right-handed neutrino mixing in the seesaw.

\end{itemize}

\subsection{Dark matter: whodunit?}

We are now ready to see what the dark matter is. It is clear from the
above discussion that none of the neutralinos/sneutrinos can be the
dark matter. The size of neutrino mass forces RPV couplings to be 
non-negligible and thus forces the neutralinos/sneutrinos to decay 
well within a second.
It is worth commenting that the limit on the
lightest neutralino/sneutrino lifetime from Big Bang nucleosynthesis
is automatically satisfied.

This leaves the gravitino as the only viable candidate for dark
matter, since its interactions are suppressed by the Planck scale on
top of the neutrino mass suppression. In what follows, we pursue
this possibility and study the consequences of gravitino being the
lightest supersymmetric particle (LSP).

The supergravity interaction~\cite{Wess:1992cp} relevant for
gravitino decay is
\begin{eqnarray}
\mathcal{L} &=& - \frac{1}{\sqrt{2} M_{\rm Pl}} \left[ \bar \chi_L
\gamma^\mu \gamma^\nu D_\nu \phi - \frac{i}{4\sqrt{2}} \bar
\lambda^a \gamma^\mu \sigma^{\nu\rho} F^a_{\nu\rho} \right] \psi_\mu
+ {\rm h.c.} \ ,
\end{eqnarray}
where $\psi_\mu$ is the gravitino field, $(\chi_L, \phi)$ and
$(\lambda^a, F^{a\mu\nu})$ belong to chiral and vector multiplet in
the MSSM, respectively. The essential feature of this interaction is
that gravitino couples to the supercurrent, just like graviton
couples to the energy-momentum tensor. These interactions themselves
are clearly not sufficient for gravitino to be able to decay, since they couple it
only to heavier super partners. 
Under the assumption of gravitino being LSP, it can only decay into SM particles and the
available channels are those in the Table~I. Obviously this requires
the R-parity violating interactions of Eq.~(1), which are anyway
necessary for neutrino masses and mixings.

\begin{table}[htb!]
\centerline{
\begin{tabular}{|c|c|}
\hline Gravitino as cold DM & Gravitino Decay Mode \\
\hline$m_{3/2}>m_{h^0}$ & $h^0+\nu$ \\
\hline $m_{3/2}>M_{W^\pm,Z^0}$ & $Z^0+\nu$, $W^\pm+\ell^\mp$ \\
\hline $m_{3/2}>m_q+m_{q'}$ & $q+\bar q'+\ell/\nu$ \\
\hline $m_{3/2}>m_\ell+m_{\ell'}$ & $\ell^++\ell'^-+\nu$ \\
\hline $m_{3/2}<2m_e$ & $\gamma+\nu$ \\
\hline\end{tabular}} \caption{Possible gravitino decay modes for
increasing gravitino mass. $h^0$ is the SM Higgs boson. More decay
channels are open for heavier gravitino mass.}
\end{table}

In the well-known case~\cite{Takayama:2000uz} of the largest neutrino mass being dominated by sneutrino VEV, the gravitino decay rate is
\begin{eqnarray}
\Gamma(\psi_\mu\to\gamma\nu) \simeq
\frac{1}{32\pi }
\frac{m_\nu}{ m_{\widetilde\gamma}}
\frac{m_{3/2}^3}{M_{\rm Pl}^2}
\simeq 10^{-50} {\rm GeV} \left(\frac{m_{3/2}}{5 \rm GeV}\right)^3
\left(\frac{1 \rm TeV}{m_{\widetilde\gamma}}\right) \ ,
\end{eqnarray}
where $m_\nu$ stands for the largest neutrino mass, with $0.03 {\rm eV} \lesssim m_\nu \lesssim 0.3 {\rm eV}$. The lower limit comes from 
atmospheric neutrinos and the upper limit from cosmological considerations~\cite{Amsler:2008zzb}.
This decay has a spectacular signature of monochromatic photons and
neutrinos. The non-observation of such signals at Fermi-LAT~\cite{Porter:2009sg} sets a lower limit on the lifetime of gravitino
$10^{26}$ seconds, or $\Gamma_{3/2}\lesssim10^{-50}\,$GeV. This in turn would
imply $m_{3/2}\lesssim5\,$GeV.

\subsection{PAMELA  and Fermi-LAT versus MSSM: setting the stage}

The striking observation of the new cosmic-ray experiments is the
substantial excess of the positrons/electrons in the range $10-100\,$GeV for PAMELA 
and up to TeV for Fermi-LAT. If confirmed, this data
could be explained by astrophysical sources~\cite{Blasi:2009hv}, e.g., pulsars~\cite{Hooper:2008kg}. On the other hand, it is also possible that
these phenomena are due to the decay or annihilation of dark matter
particles. If the latter were to be true, what would it imply for
the MSSM? In what follows, we discuss the answer to this important
question, by pursuing the scenario of gravitino dark matter which 
immediately requires its mass to be above a few hundred GeV or so. 
This then clearly eliminates the possibility of sneutrino VEV, discussed above, being
a dominant source of neutrino mass. Another
important information from PAMELA is the simultaneous lack of
antiprotons excess. {\it Whoever the culprit is, it must be
leptophilic.} This uniquely select the $\lambda$-term in Eq.~(1) as
the main source of gravitino decay~\cite{Bomark:2009zm}. 
\medskip

The usual problem of the MSSM, when it comes to discussing and
presenting the results is the proliferation of its parameters. In 
order to ease the reader's pain, we discuss first a
simplified situation where $\lambda'$ coupling is simply set to
zero.   
The result of heavy sleptons will be shown to remain valid in the general case 
presented in the next sub-section.  Furthermore, for the sake of illustration 
and simplicity, we take an imaginary situation of a single $\lambda$ coupling 
(say $\lambda= \lambda_{133}$, the corresponding 
$[(m^2_{\widetilde{\ell}})_{\rm LR}]_{33}\ne 0$,
 and all the other elements vanishing).
We will comment on the general case in Subsection~II C.

In this case the neutrino mass is given by
\begin{eqnarray}\label{neu}
m_\nu \simeq \displaystyle 
\displaystyle\frac{\lambda^2 (m_{\widetilde \ell}^2)_{LR}
m_\tau}{16\pi^2 m_{\widetilde \ell}^2} \ .
\end{eqnarray}

\medskip

The crucial thing to notice is that the gravitino decay can result
only from dimension 6 or 7 effective interactions. The essential
point is that in supergravity the gravitino is coupled to the
supercurrent which is diagonalized simultaneously with the K\"ahler
potential. These effective operators are obtained by integrating the
sfermions out, and as such are necessarily suppressed by the
sfermion masses \footnote[1]{Here we disagree with the
Ref.~\cite{Lola:2007rw}. This leads also to the different power of the gravitino mass dependence 
of the two-body decay. See the Appendix for more details.}. The
simple physical reasoning is confirmed by explicit computations
which we leave for the Appendix.

Now we list the main two- and three-body decay modes that are
induced by the leptophilic $\lambda$ coupling. The two-body decay
rates are computed in the Appendix with the following results (we
discuss the photon modes separately)
\begin{eqnarray}\label{g2}
\Gamma_2(\psi_\mu\to W^\pm\ell^\mp) &\simeq&  \frac{g^2 \lambda^2}{18432 \pi^5}
\frac{(m_{\widetilde \ell}^2)_{LR}^2 }{
 m_{\widetilde \ell}^4} \frac{m_{3/2}^3}{M_{\rm
Pl}^2 },  \\
\Gamma_2(\psi_\mu\to Z^0\nu) &\simeq& \frac{1}{2\cos^2{\theta_W}} 
 \,\Gamma_2(\psi_\mu\to W^\pm\ell^\mp)\ , \nonumber \\
\Gamma_2(\psi_\mu\to h^0\nu) &\simeq& \frac{m_{3/2}^2}{864 M_W^2} 
\Gamma_2(\psi_\mu\to W^\pm\ell^\mp) \ , \nonumber 
\end{eqnarray}
where $g$ is the $SU(2)_L$ gauge coupling and
$\theta_W$ is the weak mixing angle. 
In order to explain the PAMELA data, gravitino must weigh more than about 300\,GeV, which allows us to safely ignore the
final state masses.
The partial decay width
$\Gamma_2(\psi_\mu\to h^0\nu)$ is somewhat suppressed compared to $\Gamma_2(\psi_\mu\to W^\pm\ell^\mp)$
or $\Gamma_2(\psi_\mu\to Z^0\nu)$, for $m_{3/2}\lesssim\,$TeV. Hereafter, we will call $\Gamma_2$ the sum of the three partial decay rates in Eq.~(\ref{g2}).

Notice the leptophilic nature of the $\lambda$ coupling is not
sufficient to suppress the antiprotons, since $W$ and $Z$ bosons
decay preferentially into hadrons. These decays must be suppressed, as much as the decays
induced by $\lambda'$ (see below).

The three-body decay is given by\footnote[2]{Notice miraculously the same 
coefficient as in the two-body decay.}~\cite{moreau}
\begin{eqnarray}\label{g3}
\Gamma_3 (\psi_\mu\to\ell^+\ell^-\nu) &=&
 \frac{\lambda^2}{18432\pi^3} 
 \frac{m_{3/2}^4}{m_{\widetilde \ell}^4} 
 \frac{m_{3/2}^3}{M_{\rm Pl}^2} \ .
\end{eqnarray}
Once again, according to PAMELA and Fermi-LAT, this decay channel
must be the dominant one. We can see, at least qualitatively, that either 
the sleptons must be heavy or $\lambda$ is forced to be small; however a small
$\lambda$ suppresses neutrino mass. We quantify this now, with the result of 
slepton mass lying in the $10^2 - 10^6\,$TeV.

We list here our criteria to fit neutrino mass and PAMELA and/or Fermi-LAT with gravitino decays,
while keeping the theory perturbative.
\begin{eqnarray}
\label{numassbound}
\begin{array}{cl}
0.03\, {\rm eV}\lesssim m_\nu\lesssim 0.3\,{\rm eV}&(\nu\,{\rm mass}) \ ,\\
\label{gamma3bound}
10^{-51}\,{\rm GeV}\lesssim \Gamma_3\lesssim10^{-49}\,{\rm GeV}&(\mbox{PAMELA/Fermi-LAT}) \ ,\\
\label{gamma2bound}
\Gamma_2\lesssim\Gamma_3/10&({\rm leptophilic\,DM}) \ ,\\
\label{lambdabound}
\lambda^2\lesssim4\pi&({\rm perturbativity\,bound}) \ .
\end{array}
\end{eqnarray}
In view of the $\lambda$ induced three-body decay being dominant, we may safely take hereafter 
for the total gravitino decay rate: $\Gamma_{3/2} \simeq \Gamma_3$. The reader should not be 
confused by our notation since we will freely interchange the above decay rates when we speak of gravitino decay.

It is a simple exercise to derive a lower bound on the slepton masses using Eqs.~(\ref{neu}) -- (\ref{g3}) and 
our criteria Eq.~(\ref{numassbound})
\begin{equation}
\label{mtildelessm32}
m_{\widetilde \ell} \gtrsim  600\,{\rm TeV} 
\left(\frac{m_{3/2}}{400\,{\rm GeV}}\right)^{5/2} 
\left(\frac{m_\nu}{0.1{ \rm eV}}\right)^{1/2}
\left(\frac{\Gamma_3}{10^{-49}{\rm GeV}}\right)^{-1/2}   \ .
\end{equation}
The above formula is obtained by omitting the Higgs channel in the two-body gravitino decay,
which is a good approximation for gravitino mass roughly below TeV. Even for gravitino mass as large as
around 3\,TeV, which we take later for fitting Fermi-LAT, the correction will be only about 20\%.


We normalize $\Gamma_3$ to its largest value in order to guarantee the lower limit on 
the slepton masses.
From Eq.~(\ref{g3}), this also implies a lower bound on $\lambda$.
 At the same time, there is an upper limit on $\lambda$ from the requirement of perturbativity,
 which when needed, we take as $\lambda^2 \lesssim 4 \pi$.
  To be as complete as possible, we leave it free in the formulas below.
In this manner, one can also obtain an upper bound 
on the slepton masses from the three-body decay rate Eq.~(\ref{g3}),
\begin{equation}
\label{mtildelessm33}
m_{\widetilde \ell} \lesssim   10^4\,{\rm TeV} 
\left(\frac{\lambda_{\rm max}^2}{4 \pi}\right)^{1/4}
\left(\frac{m_{3/2}}{400\,{\rm GeV}}\right)^{7/4}
\left(\frac{\Gamma_3}{10^{-51}{\rm GeV}}\right)^{-1/4}\ .
\end{equation}
Since now we are interested in the upper limit on slepton masses, we must clearly 
normalize $\Gamma_3$ to its lowest allowed value for the sake of PAMELA.

The result of the above limits turns out to be a relatively small range of slepton masses, 
completely out of LHC reach. Notice that the region gets narrower as the gravitino 
mass increases. 

Similarly, one can obtain an upper bound on the gravitino mass which depends only on its decay rate
and the neutrino mass
\begin{equation}
\label{m32pamela}
\left(\frac{m_{3/2}}{3.0\,{\rm TeV}}\right) \left[ 0.5 + 0.5 \left( \frac{m_{3/2}}{3.0\,{\rm TeV}} \right)^2 \right]^{1/3} \lesssim \left(\frac{\lambda^2}{4 \pi}\right)^{1/3}
 \left( \frac{\Gamma_3}{10^{-49}\rm GeV} \right)^{1/3} \left( \frac{m_\nu}{0.1\rm eV} \right)^{-2/3} \ .
\end{equation}
The high-energy data of Fermi-LAT require the gravitino mass bigger than 2 TeV. We find below that the
three-body gravitino decay fit of these data works optimally with $m_{3/2}=3.3\,$TeV. This points immediately
to the necessarily large $\lambda$ coupling and the possible tension with perturbativity. 
For example
\begin{equation}\label{pert}
\lambda^2/(4\pi)\gtrsim\{0.14,\,1.5,\,14\}\ \ {\rm for}\ \  m_\nu=\{0.03,\,0.1,\,0.3\}\,\rm eV \ .
\end{equation}
Clearly, the perturbativity requirement favors hierarchical neutrino mass spectrum, and the degenerate
spectrum could actually invalidate the gravitino decay as an explanation of the Fermi-LAT data within the
perturbative MSSM. 
Of course, we are interested in as small values
of $m_{3/2}$ as possible for the sake of experiment. Recall that the gravitino is assumed here 
to be an LSP and heavy gravitino, although perfectly acceptable, means no supersymmetry at 
colliders such as the LHC. It is interesting that at the same time, the gravitino decay solution to the 
electron/positron excesses in the cosmic rays favors the light gravitino.
 
 From Eq.~(\ref{pert}), one obtains the upper bound on gravitino mass $m_{3/2} \lesssim 6.6\,$TeV by requiring $\lambda^2/(4 \pi) \lesssim 1$ and by
 choosing $m_\nu = 0.03\,$eV. In turn, one gets the upper limit on the slepton mass $m_{\widetilde \ell} \lesssim 10^6\,$TeV form Eq.~(\ref{mtildelessm33}).

\begin{itemize}
\item Let us turn our attention to PAMELA first and ignore Fermi-LAT for the time
being. We use the decay modes $\psi_\mu\to\ell^+\ell^-\nu$ to fit
the PAMELA observation of positron spectrum. 
In the left-panel of Fig.~1, we show the allowed region in the
$\lambda-m_{\widetilde\ell}$ plane. By
choosing the gravitino mass to be 400 GeV, the PAMELA data confine
the slepton masses to the range from 560 TeV to $1.7\times10^4$ TeV. This agrees well with our 
previous estimates in Eqs.~(\ref{mtildelessm32}) and (\ref{mtildelessm33}). 

\begin{figure}[h!]
\begin{center}
\includegraphics[width=7cm]{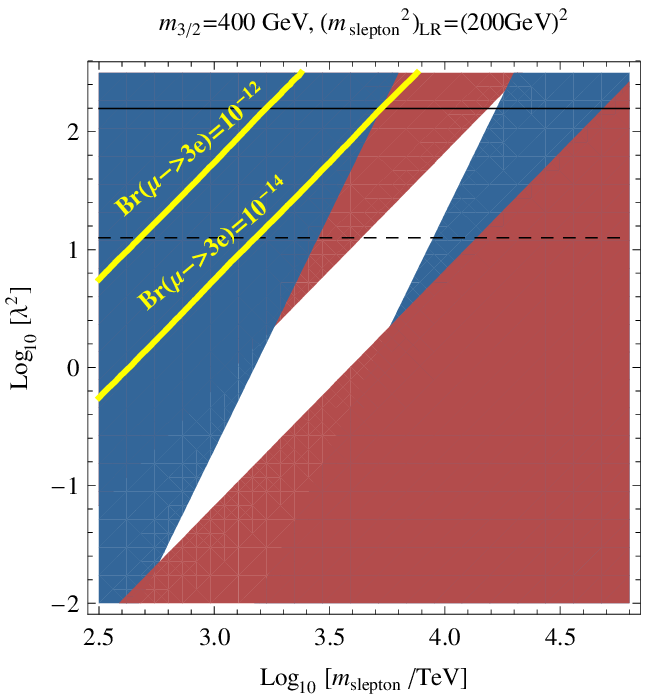}
\includegraphics[width=7.2cm]{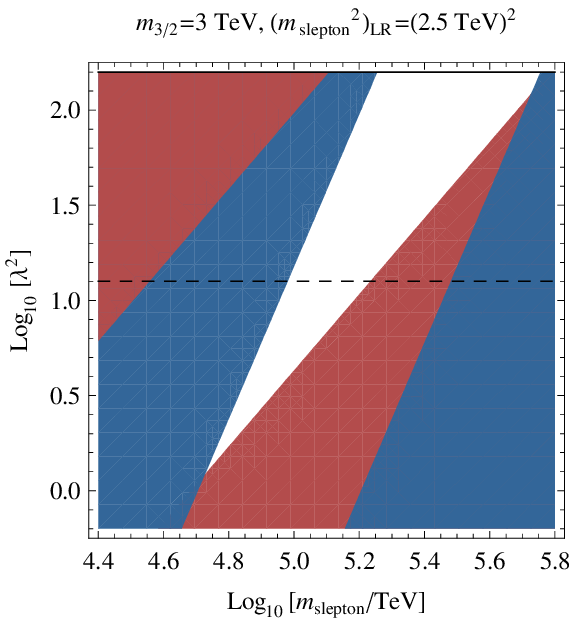}
\caption{(colored online) Left-panel: Allowed region in
$\lambda^2-m_{\widetilde \ell}$ parameter space for $m_{3/2}=
400\,$GeV in order to explain PAMELA. All colored regions are excluded. 
 Assuming that the $\lambda$'s responsible for gravitino decay are the same as
the ones causing LFV $\mu \to 3 e$ decay, the yellow lines represent contours of ${\rm Br}(\mu\to3e)$
(the present experimental upper bound is $10^{-12}$). 
Right-panel: Allowed region in
$\lambda^2-m_{\widetilde \ell}$ parameter space for $m_{3/2}= 3\,$TeV 
in order to explain Fermi-LAT. All colored regions are excluded. In both figures, the constraints
are $10^{-51} {\rm GeV} < \Gamma_3 < 10^{-49}\,{\rm
GeV}$ (blue) and
$0.03 {\rm eV} <m_\nu < 0.3\,{\rm eV}$ (red). The usual perturbativity regions $\lambda^2 \leq 4 \pi $ lie below the horizontal dashed lines. 
We also show the  cut-off value $\lambda^2 =16 \pi^2$ by the horizontal full lines (coincides with the top of the figure in the right panel).
\label{lambdamslepton}}
\end{center}
\end{figure}

If the neutrino spectrum were to be determined, or if at least we were to know the largest
neutrino mass, the slepton masses would be directly correlated with the $\lambda$ coupling.
In the Fig.~2 we give this dependence with $m_{3/2} = 400 {\rm GeV}$ value for PAMELA, 
indicated by a blue line. The different color shaded regions below the red band are forbidden  
for different values of neutrino mass, indicated inside the corresponding regions.  

\begin{figure}[htb]
\begin{center}
\includegraphics[width=9cm]{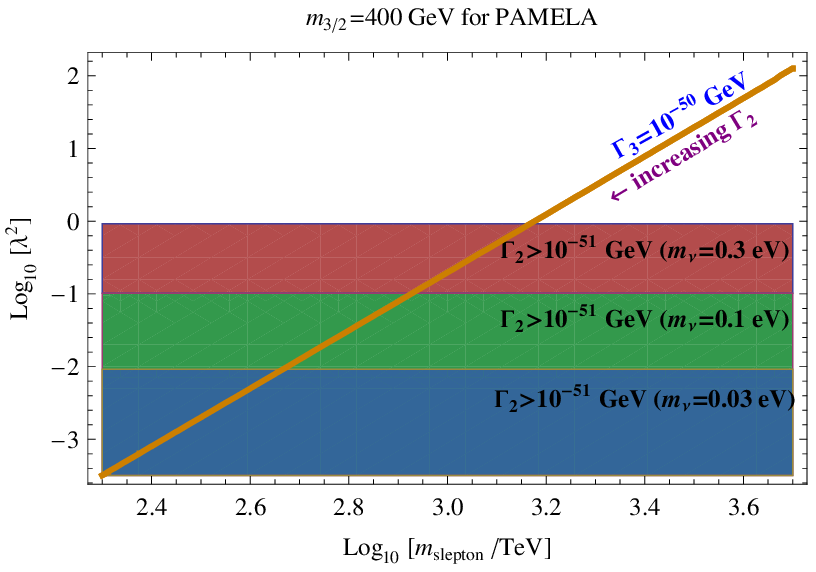}
\caption{(colored online) $\lambda$ -- slepton mass dependence for 
$m_{3/2}=400$ GeV and $\tau_{3/2}=0.65\times10^{26}\,$sec ($\Gamma_{3/2}=10^{-50}\,$GeV), for different values of neutrino mass. The three colored areas denote the forbidden 
parameter spaces.}
\end{center}
\end{figure}

\item
Our fit for PAMELA data is shown in Fig.~3. We choose $m_{3/2}=400\,$GeV and
$\tau_{3/2} = 2.3\times 10^{26}\,$sec  ($\Gamma_{3/2}=0.3\times10^{-50}\,$GeV).
The best fit fixes the ratio $m_{\widetilde\ell}^2/\lambda \simeq1.3\times10^7\,$TeV$^2$.
The leptons in final states of gravitino decay
are taken to be electrons\footnote[3]{This fit is similar to the one in Ref.~\cite{Bomark:2009zm}, 
who however assumed the slepton mass to be in the TeV region. The issue of course is the 
implication for the neutrino mass, which pushes the sleptons to be much heavier.}.
We have adopted the solution to the positron transportation equation from Ref.~\cite{Ibarra:2008qg}, 
and the parametrizations of the electron/positron background of Ref.~\cite{Baltz:1998xv}~\cite{Baltz:2001ir}.


\begin{figure}[t!]
\begin{center}
\includegraphics[width=9cm]{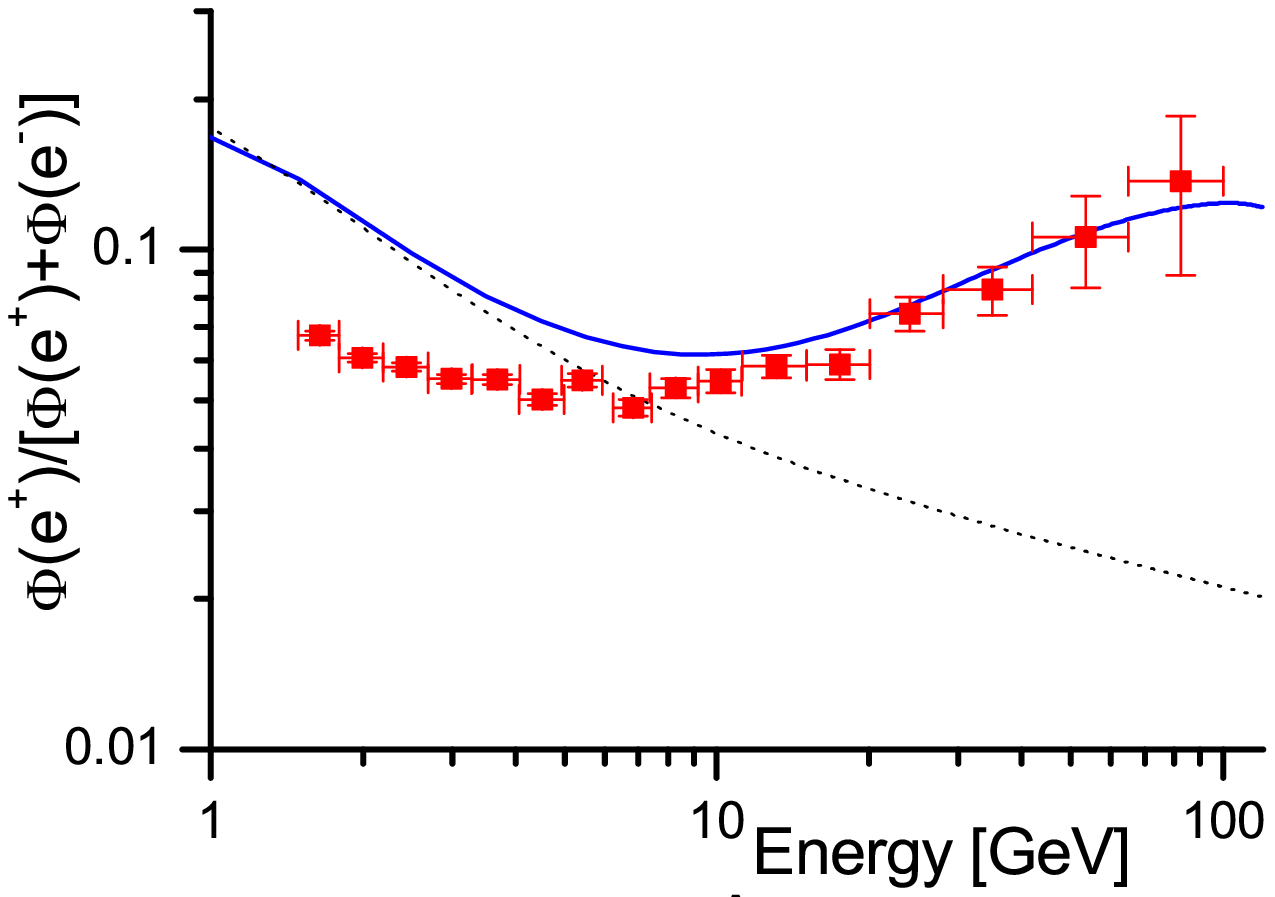}
\caption{(colored online) The fit for the PAMELA positron excess using the three-body gravitino decay;
 with $m_{3/2}=400$ GeV, $\tau_{3/2} = 2.3\times 10^{26}\,$sec ($\Gamma_{3/2}=0.3\times10^{-50}\,$GeV)
and neutrino mass 0.2\,eV. The red dots represent the data, the blue solid curve is our fit whereas the dotted black curve
is the expected theoretical background. 
The discrepancy between the data and the theoretical prediction at low energies is commonly explained by the solar modulations.}  
\end{center}
\end{figure}

\item It is also useful to look into the available $m_{\widetilde \ell}-m_{3/2}$ plane as Fig.~4, 
instead of the $\lambda-m_{\widetilde \ell}$ plane used so far. Notice the upper bound on the gravitino mass
as discussed above.

\begin{figure}[htb]
\begin{center}
\includegraphics[width=10cm]{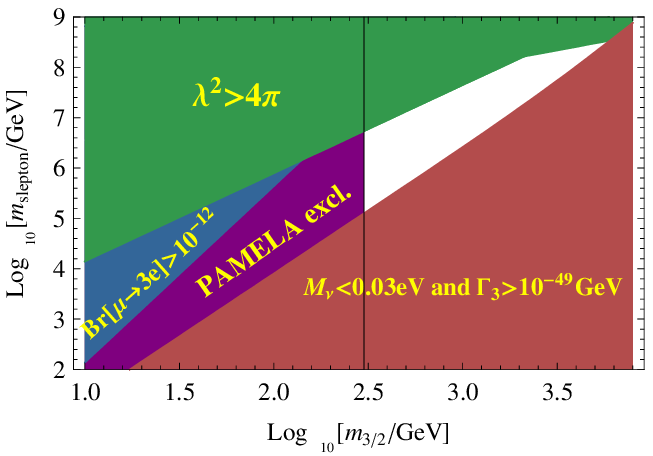}
\caption{(colored online) The allowed region (white) in the $m_{\widetilde \ell}-m_{3/2}$ parameter space. The red and green regions are 
excluded by Eq.~(\ref{mtildelessm32}) and Eq.~(\ref{mtildelessm33}), respectively, due to PAMELA. 
The region where the gravitino is lighter than 300 GeV is also ruled out. One can see that the LFV process ${\rm Br}(\mu\to3e)$ 
is safely small for the allowed region. 
\label{mtildam32}}
\end{center}
\end{figure}

\begin{figure}[htb]
\begin{center}
\includegraphics[width=8cm]{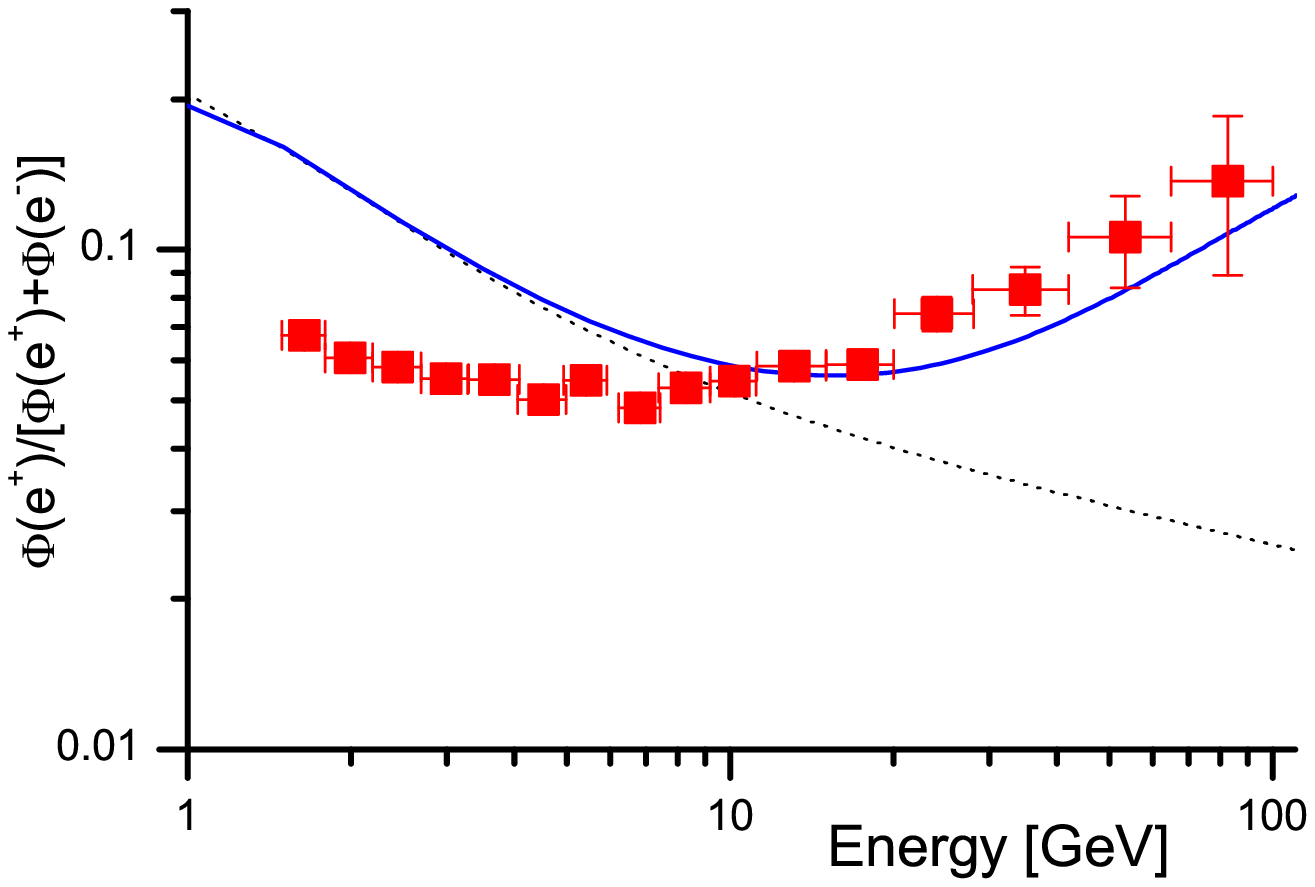}
\includegraphics[width=8cm]{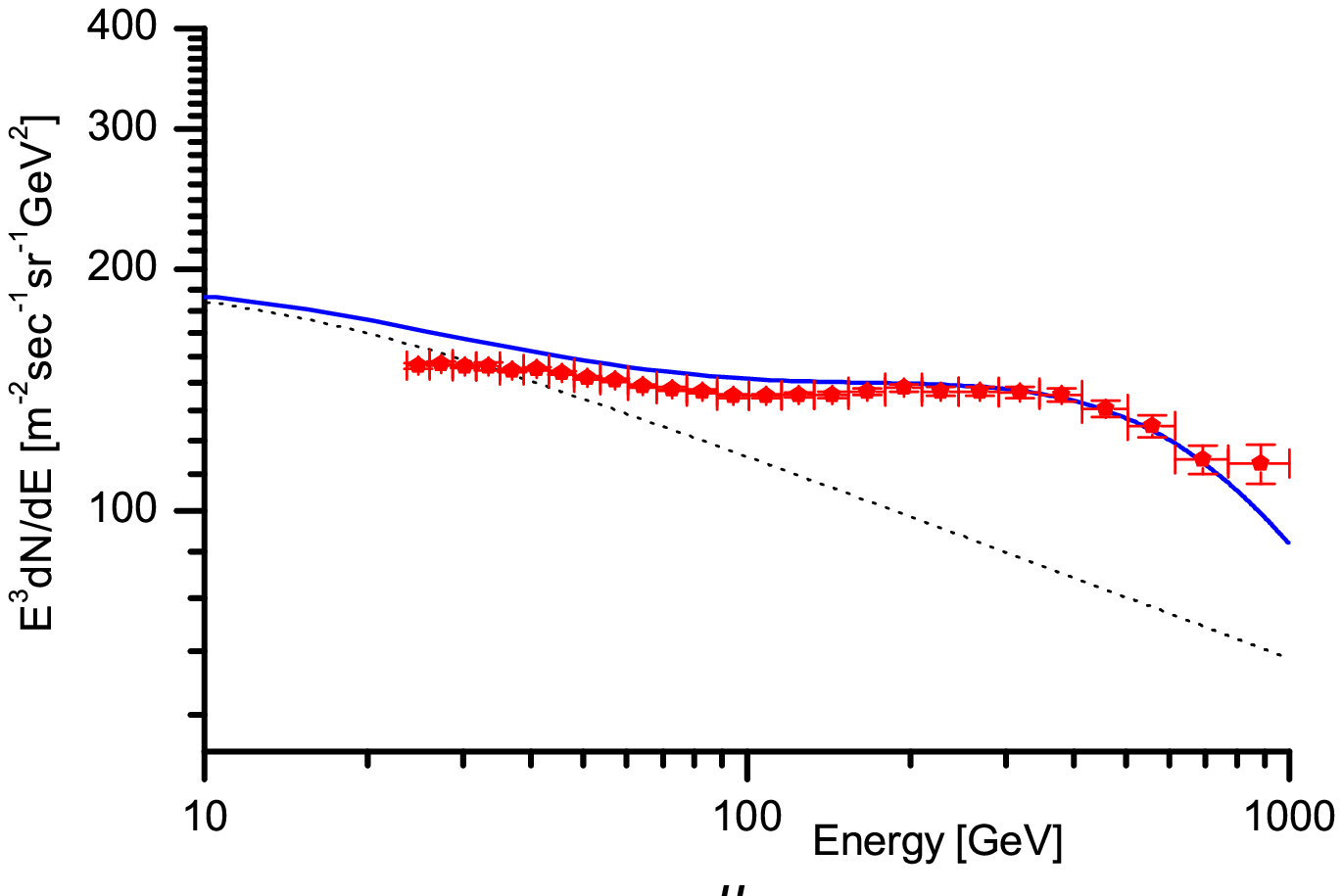}
\caption{(colored online) The three-body gravitino decay is used to fit both PAMELA and Fermi-LAT
electron/positron, with $m_{3/2}=3.3$ TeV, $\tau_{3/2} = 5\times 
10^{25}$sec ($\Gamma_{3/2} = 1.4 \times 10^{-50}\,$GeV) and neutrino mass 0.03 eV.  
The red dots represent the data, the blue solid curve is our fit whereas the dotted black curve
is the expected theoretical background. To guarantee the soft spectrum, we demand only ten
percent of the final lepton states are electrons.} 
\end{center}
\end{figure}

\item Next, we consider the possibility of interpreting both PAMELA and
Fermi-LAT simultaneously. For decaying gravitino dark matter,
Fermi-LAT data generally requires its mass to be greater than $2-3\,$TeV. Taking
into account the same constraints as the PAMELA case discussed
before, we find the sleptons even heavier, more than
$4.5\times10^4$ TeV as shown in the right-panel of Fig.~1. With such
heavy sleptons, all the LFV constraints become irrelevant. 

In Fig.~5, we fit both PAMELA positron and Fermi-LAT positron/electron
excesses using a 3.3 TeV gravitino mass. We take the gravitino lifetime
$\tau_{3/2} = 5\times 
10^{25}$sec ($\Gamma_{3/2} = 1.4 \times 10^{-50}\,$GeV). The best fit fixes now the ratio $m_{\widetilde \ell}^2/\lambda \simeq 10^{10}\,$TeV$^2$. To guarantee the soft electron/positron
spectrum observed, we demand only 10\% of the final lepton states from
primary gravitino decay are electrons~\cite{Papucci:2009gd}. We took the heaviest neutrino mass of 0.03 eV,
i.e. the hierarchical neutrino spectrum, while the inequality $\Gamma_2\lesssim\Gamma_3/10$ is saturated.
Taking the neutrino mass degenerate endangers the perturbativity as we discussed at length.
Notice that in this simple picture, the more precise experiments with a precision
improved by an order of magnitude, would have to see the excess of antiprotons too.

\end{itemize}

\subsection{Towards the general case}

\subsubsection{Splitting $m_{\widetilde L}$ and $m_{\widetilde e^c}$}

For simplicity, we took above $m_{\widetilde L}\simeq m_{\widetilde e^c}$ and we learned that the two sleptons
must be quite heavy. Let us relax this assumption and denote by $m_L$ and $m_H$ the small and large
slepton masses (we treat    
the left- and right- slepton mixing mass $(m_{\widetilde \ell}^2)_{LR}$ as a perturbation).

Since we know the result when the masses are similar, it is worthwhile investigating only the case when they are
widely split, $m_H\gg m_L$. It is easy to see that all the above formulas go through, modulo the following changes.
\begin{eqnarray}
m_{\widetilde\ell}^2 &\to& m_L^2 \nonumber \\
(m_{\widetilde \ell}^2)_{LR} &\to& r (m_{\widetilde \ell}^2)_{LR} 
\end{eqnarray}
where $r\simeq (m_L/m_H)^2 \ln (m_L/m_H)^2$. In turn, the bounds on $m_{\widetilde \ell}$ become the bounds on $m_L$.

The lower limit on $r$ emerges from the requirement that neutrino mass be large enough, $m_{\nu}\gtrsim0.03\,$eV.
Barring cancellations, one gets $r\gtrsim10^{-7}$. In other words, one of the two sleptons can be 3-4 orders of
magnitude heavier than the other. We have assumed here that the lighter slepton is not fine-tuned to decouple;
we discuss this possibility in Subsection~\ref{C4}.

\subsubsection{$\lambda$ with family indices}

Before discussing the general case, let us comment on the multi-generation situation 
with $\lambda$ couplings. This is quite involved and basically impossible to have a 
simple conclusion. As such, it is beyond the scope of this paper. Up to now, we have
imagined a situation with a single coupling, an extreme simplification. Another extreme
possibility is the situation when all elements in $\lambda$ matrix
are equal in magnitude, i.e., $\lambda_{ijk}=\lambda$ for all the nine complex elements
in the physical basis. 

The neutrino mass matrix and the two-body gravitino decays depend also directly on the mixings
between left and right sleptons, and in principle, the end result may be similar to the simplified 
case with a single coupling. The three-body decay is more sensitive to the generation
structure, because the final states contribute democratically. Assuming for example, universal 
soft masses for the sleptons, one simply obtains an enhancement of factor of nine in the total decay 
rate. Due to the quartic dependence on the slepton masses, even this extreme case has a minor impact
on the conclusion for the single $\lambda$ case.

Clearly, barring cancellations, a generic case lies between
these two opposite cases. In short, our result of heavy sleptons goes unchanged,
except for a possibility of fine-tuning one of the slepton masses which we discuss 
in the the end of this section.
 
\subsubsection{$\lambda'$ and the fate of squarks}

In the previous subsections, we have simplified the discussion by choosing $\lambda$ couplings 
to be the only non-vanishing ones. The motivation was the leptophilic nature of PAMELA and
Fermi-LAT data, which prefers the $\lambda'$ couplings to be relatively small. 
\begin{eqnarray}
\lambda' \lesssim \lambda'^{\rm max} = 4.5\times10^{-8} \left( \frac{m_{\widetilde q}}{1\,\rm TeV} \right)^{2} \left( \frac{m_{3/2}}{400\,\rm GeV} \right)^{-7/2}\  ,
\end{eqnarray}
where we used the bound $\Gamma_3  \lesssim 10^{-49}\,$GeV employed throughout this paper, and we demand that 
the hadronic gravitino partial decay rate be smaller by an order of magnitude.

However, they 
may still contribute to neutrino mass, in which case the main result remains unchanged: the sleptons must be heavy as before. If $\lambda'$ couplings are non negligible, the squarks also end up
being heavy, for the same reason as Eq.~(\ref{mtildelessm32}). 

Needless to say, the limit on squark masses depends on the degree of the $\lambda'$ contribution
to the neutrino mass matrix. On pure phenomenological grounds, squark masses
are completely free.


\subsubsection{On a possibility of having a light slepton(s)}\label{C4}
     
So far in the discussion, we have suppressed the flavor indices or assumed democratic matrix elements of $\lambda_{ijk}$. 
The question then arises 
whether there is a loophole in our argument, i.e., whether one can adjust the parameters 
to make one (or more) slepton light (in TeV scale) in the general multi-generation case. This could happen
for example if a slepton (or more than one) has vanishing RPV
couplings in the physical basis.  Although unlikely,
this seems to us perfectly acceptable; and in all fairness we cannot claim that all sleptons are
necessarily heavy as in the above warm-up examples.
In general, to decrease the mass of a slepton $\widetilde \ell_1$, one
must also decrease its RPV couplings $\lambda_1$ in a manner that 
the ratio $\lambda_1/m_{\widetilde \ell_1}^2$ does not get enhanced. This will guarantee the
gravitino decay rate and neutrino masses do not receive too large contribution from
$\widetilde \ell_1$. From this requirement, using the gravitino decay
rate given in Eq.~(\ref{g3}), we have the following upper bound 
\begin{eqnarray}\label{LightSlep1}
\lambda_{1}   \lesssim \lambda_1^{\rm max}=4.5\times10^{-7} \left( \frac{m_{\widetilde \ell_1}}{1\,\rm TeV} \right)^{2} \left( \frac{m_{3/2}}{400\,\rm GeV} \right)^{-7/2}\  ,
\end{eqnarray}
where we again used the bound $\Gamma_3  \lesssim 10^{-49}\,$GeV. 

The fact that the light slepton (almost) decouples from RPV ensures the
suppression of its contribution to the LFV processes through RPV.
However, this does not prove the absence of LFV,
since it can always take place through the usual soft supersymmetric
breaking, which in general arises from different mixings in the sfermion and fermion sectors.  
The reader probably dislikes this fine-tuned possibility as much as we do, and it is fair to say that one expects the sleptons
to be heavy. This was behind our statement in the introduction that PAMELA and Fermi-LAT
imply heavy sleptons, barring fine-tuned cancellations. In a sense, this could be viewed as a blessing,
since the flavor problem in supersymmetry is one of its weakest points and heavy sfermions automatically
take care of this. We interpret our results as an argument in support of moderately split supersymmetry.

    This said, we should also recall the main motivation for low energy supersymmetry: the hierarchy
    problem. A heavy slepton aligned mostly in a stau direction upsets mildly the Higgs mass naturalness,
    and so it could be appealing that this particle be light. The sleptons aligned in the smuon and selectron
    direction are clearly allowed to have masses on the order of $10^2 - 10^4\,$TeV without upsetting
    the tree-level stability of the Higgs mass.  

 \section{Phenomenological and cosmological implications}

Here we comment on salient features regarding various collider, flavour violation and cosmological phenomena
that are special to gravitino decay being behind PAMELA and/or Fermi-LAT. 

\subsection{LHC signatures}

As we have shown, both PAMELA and Fermi-LAT can be explained via gravitino decay, as long as the gravtino is
heavier than about several TeV. In this case, clearly no collider physics will emerge in near future.

Phenomenologically, a more interesting scenario is to have gravitino behind PAMELA only, in which case, the gravitino mass can be as low as several hundred GeV. Except for heavy sleptons, if gravitino were to be this light, the rest of superpartners could be observable at the LHC. It is then important to study possible signatures of the NLSP. There are a number of NLSP candidates that we discuss in the following.

\medskip
\paragraph*{Gaugino as the NSLP.} First, we consider the case where the NLSP is mainly of wino/bino--type.  In the presence of the R-parity violating terms, the NLSP
decays to three leptons at tree level
very much like gravitino decay. When three-body decay channel dominates over that of the two-body, one can express the decay width in terms of gravitino
decay rate as follows
\begin{eqnarray}
{\Gamma}_{_{\rm NLSP}}({\widetilde\chi_1^0}\to \ell^+\ell^-\nu) = \frac{g^2 \lambda^2}{3072\pi^3} \frac{m_{_{\rm NLSP}}^5}{m_{\widetilde \ell}^4} = 
\frac{6g^2M_{\rm Pl}^2 m_{_{\rm NLSP}}^5}{m_{3/2}^7} \Gamma_{3/2}\ .
\end{eqnarray}
The squark exchange contribution 
is similarly subdominant as for the gravitino decay. If we fit PAMELA only, with 
$m_{3/2}=400$ GeV and $\Gamma_{3/2}=10^{-50}\,$GeV, one obtains the NLSP lifetime
\begin{eqnarray}
\tau_{_{\rm NLSP}}^{\widetilde\chi_1^0} \simeq  10^{-7} {\rm sec}\,\left( \frac{m_{_{\rm NLSP}}}{600\,{\rm GeV}} \right)^{-5} \ ,
\end{eqnarray}
which corresponds to the decay length at the LHC $d_{_{\rm NLSP}}\simeq 30 \,$m for $m_{_{\rm NLSP}}=450-600\,$GeV.
We can see that the decay length of the bino/wino NLSP is generically rather long. Still, once produced at the LHC,  
a sizable amount of these particles appears to 
decay inside the detector~\cite{Ishiwata:2008tp}, producing highly-ionizing charged tracks 
(if the NLSP is charged wino, it has similar decay rate) with multi-lepton final states. 

Although the above expression is written with a single, generic coupling, it is valid in a general
situation with arbitrary RPV couplings, for the same couplings enter in the LSP (gravitino) and NLSP
three-body decays. It is a solid prediction of the MSSM being behind PAMELA.

In the less-likely scenario that the gluino be the NLSP, it would have to decay through $\lambda'$, by 
the analogy with the above wino/bino decay. The point is that the gluino must decay before BBN, and the Planck
suppressed decay into a gravitino is too slow. This leads immediately to a lower limit on $\lambda'$ 
\begin{eqnarray}
\lambda' \gtrsim  \lambda'^{\rm min} = 1.28\times 10^{-11} \left( \frac{m_{\widetilde q}}{1\,\rm TeV} \right)^{2} 
\left( \frac{m_{_{\rm NLSP}}}{600\,\rm GeV} \right)^{-7/2}\  .
\end{eqnarray}

\medskip

\paragraph*{Light slepton as the NLSP.} As discussed in the previous section, there is also the possibility of a light slepton.
However, this does not upset the above prediction if the NLSP is still gaugino type, since the light
slepton has to (almost) decouple from RPV coupling.

It is interesting to consider the possibility of the light slepton being the NLSP.
From Eq.~(\ref{LightSlep1}), we obtain an upper bound on its two-body decay rate 
\begin{eqnarray}
{\Gamma}_{_{\rm NLSP}}(\widetilde\ell_1\to\ell_j\ell_k) = \frac{\lambda_1^2 m_{_{\rm NLSP}}}{8\pi}  \lesssim 6 \times 10^{-13}\,{\rm GeV} 
\left( \frac{m_{_{\rm NLSP}}}{600\, \rm GeV} \right)^{5} 
\left(\frac{m_{3/2}}{400\,\rm GeV}\right)^{-7} \ .
\end{eqnarray}
 At the LHC, the slepton NLSP has to be pair produced.
For the charged one, there would be a displaced vertex and/or a heavily-ionizing charged track (for
sufficient small RPV couplings) with di-leptons plus 
missing energy, whereas for sneutrino NLSP, one would see two charged leptons in the final states.

\begin{figure}[htb]
\begin{center}
\includegraphics[width=7cm]{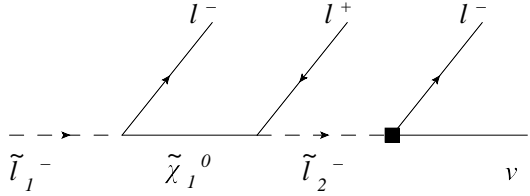}
\caption{A representative four-lepton decay mode for a slepton NLSP. The black box represents RPV coupling.}\label{slepton}
\end{center}
\end{figure}

We also consider an extreme case when the slepton NLSP completely decouples from direct RPV violation in the physical basis, i.e., $\lambda_1=0$.
Then it can only decay to four-lepton final states through a virtual gaugino $\widetilde \chi_1$, 
and a virtual heavy slepton $\widetilde\ell_2$ which possesses RPV couplings, as shown in Fig.~{\ref{slepton}}. We estimate its decay rate
\begin{eqnarray}
{\Gamma}_{_{\rm NLSP}}(\widetilde\ell_1^-\to\ell^-\ell^+\ell^-\nu) \simeq \frac{g^4 \lambda^2}{10^5\pi^5} \frac{m_{_{\rm NLSP}}^7}{m_{\widetilde\chi_1}^2 m_{\widetilde \ell}^4} 
\simeq 10^{-3} \frac{M_{\rm Pl}^2}{m_{\widetilde\chi_1}^2} \left( \frac{m_{_{\rm NLSP}}}{m_{3/2}} \right)^7 \Gamma_{3/2}\ .
\end{eqnarray}
Choosing $m_{3/2}=400$ GeV and $\Gamma_{3/2}=10^{-50}\,$GeV, one obtains
\begin{eqnarray}
\tau_{_{\rm NLSP}}^{\widetilde\ell_1} \simeq  10^{-3} {\rm sec}\,\left( \frac{m_{_{\rm NLSP}}}{600\,{\rm GeV}} \right)^{-7} 
\left( \frac{m_{\widetilde\chi_1}}{1\,{\rm TeV}} \right)^{2} \ .
\end{eqnarray}
In this case, the NLSP would decay outside the detector.

\subsection{Flavor violation}

As we have seen, the size of the R-parity breaking couplings
(called generically $\lambda$) can be as large as $\mathcal{O}(1)$. Therefore, one needs
to examine lepton flavor violating (LFV) processes. The most
stringent constraint comes from the decay $\mu\to3e$ which is
induced by the $\lambda$ coupling at tree level. Its branching ratio
is given by
\begin{eqnarray}
{\rm B}(\mu\to3e) \simeq \left( \frac{\lambda}{g} \right)^4 \left(
\frac{M_W}{m_{\widetilde \ell}} \right)^4 \ .
\end{eqnarray}
It is worth noting that at least two non-zero elements of the $\lambda$ matrix are necessary for this decay.
Here we just assume they are equal to the one responsible for gravitino decay.
We find thus that the allowed region predicts very tiny $\mu\to3e$ branching
ratio ($\lesssim10^{-16}$), as indicated in Fig.~1. The branching ratio is clearly too small to be probed in near future.
The other LFV processes such as $\mu\to e\gamma$ and
$\mu\to e$ conversion take place only at the loop level and are even
more suppressed.

  This is certainly true if the slepton masses are roughly universal in size, a natural expectation. 
  However, as we discussed above, one (or more) sleptons can be made light by an artificial cancellation
  and thus can lead to LFV as in the conventional picture of low energy supersymmetry.

 On the other hand, the squarks can be light, in which case the smallness of quark flavor violation would
 remain as a mystery. 

\subsection{Baryogenesis}

One may ask whether our findings are in accord with baryogenesis in the MSSM. Clearly, gravitino
behind PAMELA eliminates the electro-weak baryogenesis, for then the stop is too heavy to allow for a 
necessary first-order phase transition. In the MSSM defined the way we did, there is no room for the 
leptogenesis, but then we cannot pretend to know the physics at very high energies. In any case, even
if leptogenesis takes place, or any primordial lepton number were to exist, the large L-number violating $\lambda$
couplings would definitely erase it.
 This is not a problem, since there is 
  the appealing Affleck-Dine mechanism~\cite{Affleck:1984fy} which utilizes the baryon and lepton number violating flat (or almost
   flat) directions as inflatons~\cite{Allahverdi:2006iq}.
 This then leads to a baryon and lepton number asymmetric universe. Again, it is the 
   baryon number that would survive today, which makes the $u^cd^cd^c$ direction preferable.

\section{The message to take home }

      The great virtues of the MSSM are often stressed: the stabilization of the 
    gauge hierarchy, the prediction of the unification of gauge couplings, the connection
    of the Higgs and the stop mass, the radiative Higgs mechanism. Its main problem 
  is that the parameter space of the MSSM is so big that it is really a collection of theories. 
 The truth is that we know nothing about the supersymmetry breaking
and the phenomenology of the MSSM ought to be done in its proper parameter space, unless
one wishes to give only some benchmarks when say discussing the possibilities at LHC.

    The main point of taking the MSSM seriously as a theory of neutrino mass is that 
   the gravitino is the only particle stable enough to be the dark matter candidate.
   What happens then if one
   asks that the PAMELA data be explained by the dark matter? This subject is at the heart of
   our study reported here and the results are the following.
   
   \begin{itemize}
 \item Barring fine-tunings, the sleptons are heavy, with their masses lying between 
 $500$ and $10^6\,$TeV. 
 This naturally suppresses all LFV, normally a major issue in the MSSM. With the masses close to the
 lower end, the gauge hierarchy is barely destabilized. 

 \item There is no limit whatsoever on the squark masses. If the squarks were to play, though,  an important role in generating the neutrino masses,
   the leptophilic nature of gravitino decay would imply them to be heavy.  This would explain in turn 
   a mystery of small QFV. The resulting picture could then be a moderately split supersymmetry.

   \item
       
The theory can also reproduce the Fermi-LAT data with the gravitino mass around 3 TeV,
but there is a potential conflict with the perturbativity. We find it quite interesting that the calculability
of the MSSM points towards smaller gravitino mass, which is appealing from the experimental point of view.
The heavy gravitino explanation of Fermi-LAT would completely kill the hope for the MSSM to be the theory relevant for the LHC.
\end{itemize}

\section*{Acknowledgments}

We would like to thank Shao-Long Chen, Gia Dvali and Sourov Roy for useful discussions.
T.E., D.K.G. and G.S. are grateful to Svjetlana
  Fajfer and other members of the theory group at "Jo\v{z}ef  Stefan" Institute in Ljubljana for their warm hospitality during the initial stages of this work. D.K.G. also thanks ICTP High Energy Group for the hospitality when part of this work was done. 
   The work of T.E., G.S. and Y.Z. is partially supported by the EU FP6 Marie Curie Research and Training Network "UniverseNet" (MRTN-CT-2006-035863). G.S. is thankful for the Senior Scientist Award from the Ministry of Science and Technology of Slovenia. D.K.G. acknowledges partial support from the Department of Science and Technology, India under
   grant SR/S2/HEP-12/2006.

\section*{Appendix: Effective Operators for Gravitino Decay}

In this appendix, we explicitly present our results on all relevant
operators for gravitino decays.

First, the dimension seven operators relevant for gravitino three-body
decay is
\begin{eqnarray}
\mathcal{L}_{\rm eff} &=& \frac{\lambda}{\sqrt{2} M_{\rm Pl}
m^2_{\widetilde \ell}} \left[ ( \overline L
\overleftarrow{\partial}_\nu \mathbb{P}_R \gamma^\mu \gamma^\nu
\psi_\mu ) ( \overline{L} \mathbb{P}_R e ) + ( \overline e
\overleftarrow{\partial}_\nu \mathbb{P}_L \gamma^\mu \gamma^\nu
\psi_\mu ) ( \overline{L^c} \mathbb{P}_L L ) \right] + \rm h.c. \ ,
\end{eqnarray}
where we have suppressed the flavor indices in $\lambda$-coupling as well as
the final-state leptons. The above effective interaction is obtained by integrating out the heavy sleptons exchanged,
and here it is assumed all the sleptons have universal mass, i.e., $m_{\widetilde\ell}^2\equiv m_{\widetilde L}^2 \simeq m_{\widetilde e^c}^2$.

\begin{figure}[htb]
\begin{center}
\includegraphics[width=14cm]{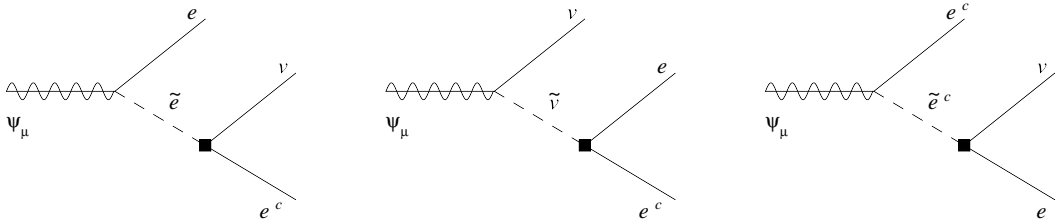}
\caption{Three-body gravitino decays via intermediate sleptons. The black boxes represent RPV couplings.}
\end{center}
\end{figure}

\begin{figure}[htb]
\begin{center}
\includegraphics[width=15cm]{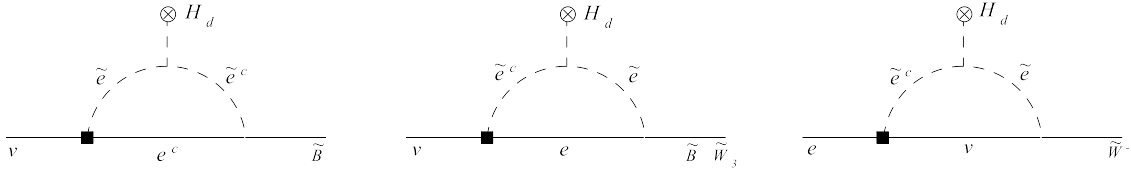}
\caption{Lepton-gaugino mixing induced by soft supersymmetry
breaking terms. The black boxes represent RPV couplings.}\label{fig7}
\end{center}
\end{figure}

\begin{figure}[t!]
\begin{center}
\includegraphics[width=4.5cm, angle=0]{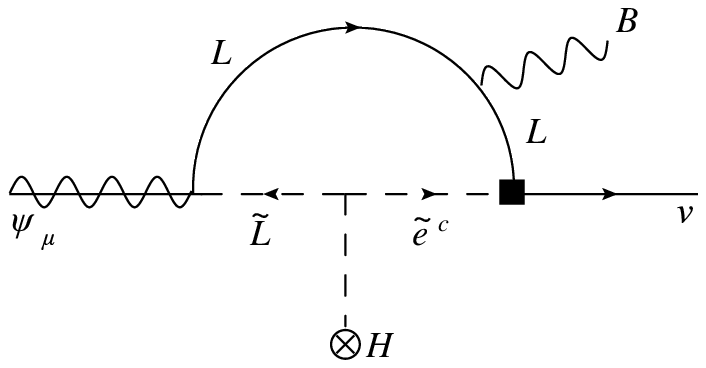}
\includegraphics[width=4.5cm, angle=0]{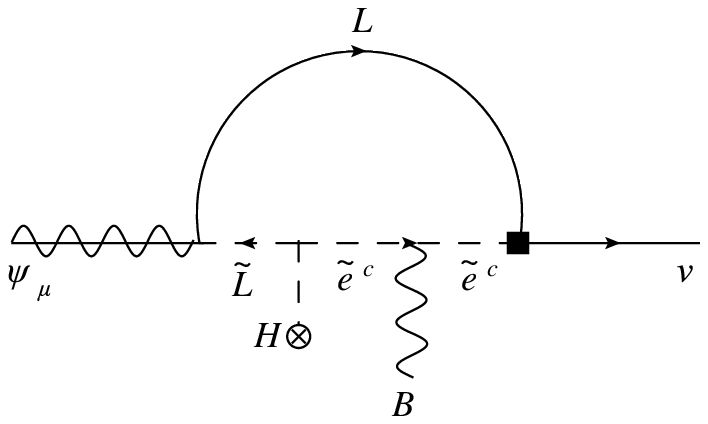}
\includegraphics[width=4.5cm, angle=0]{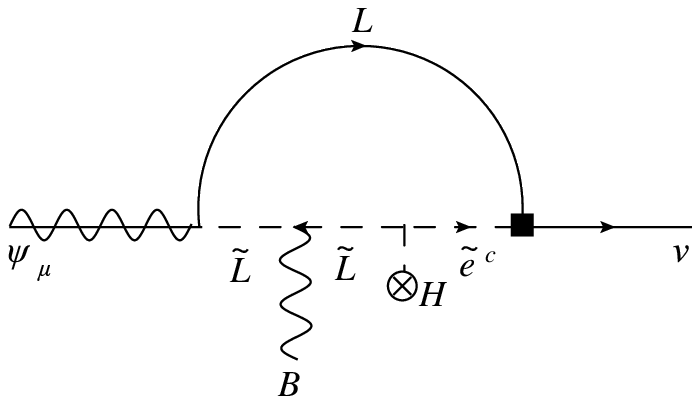}\\
\includegraphics[width=4.5cm, angle=0]{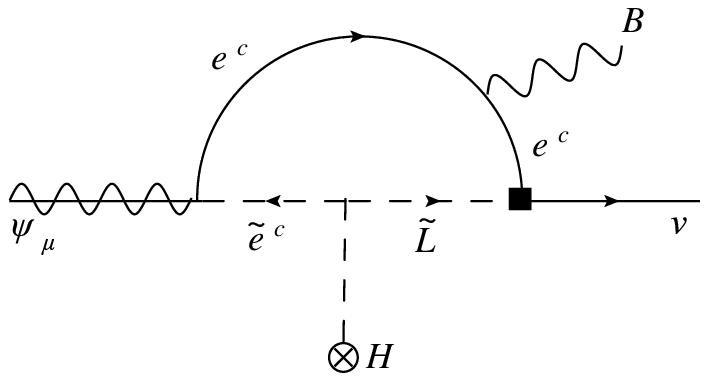}
\includegraphics[width=4.5cm, angle=0]{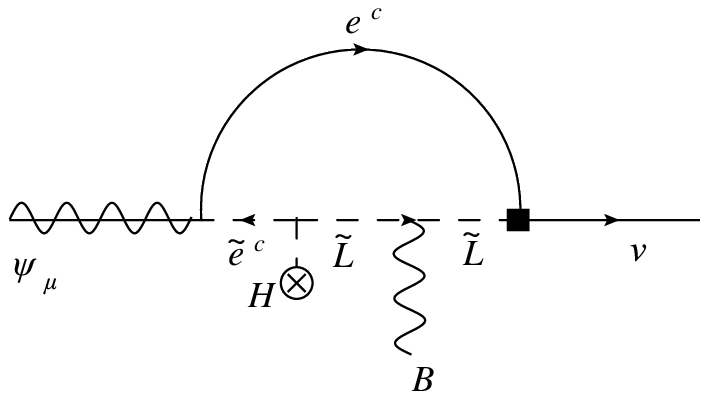}
\includegraphics[width=4.5cm, angle=0]{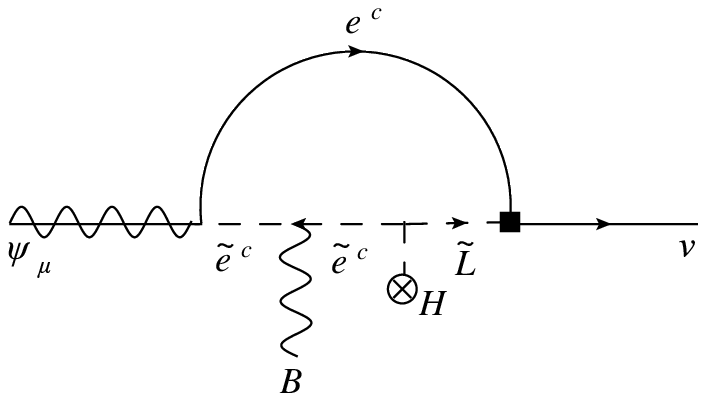}\\
\includegraphics[width=4.5cm, angle=0]{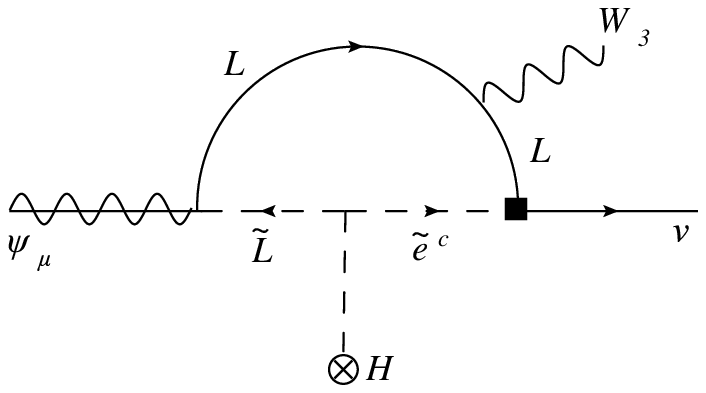}
\includegraphics[width=4.5cm, angle=0]{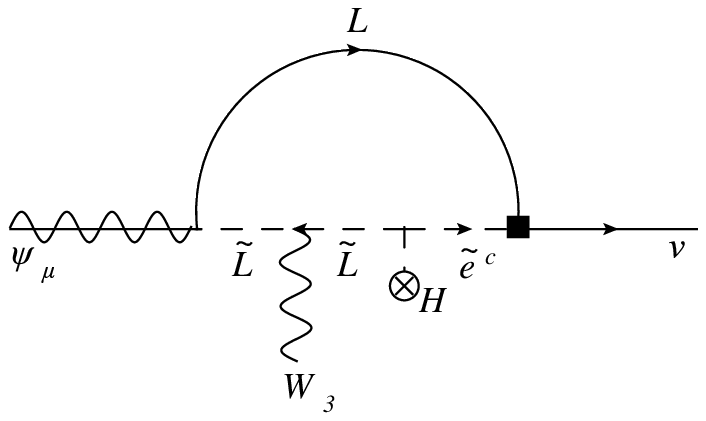}
\includegraphics[width=4.5cm, angle=0]{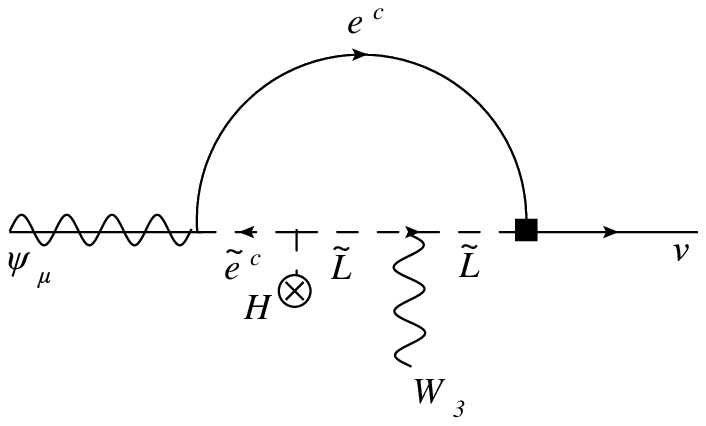}\\
\includegraphics[width=4.5cm, angle=0]{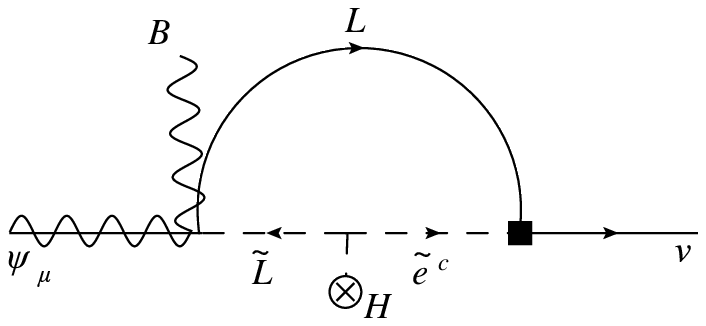}
\includegraphics[width=4.5cm, angle=0]{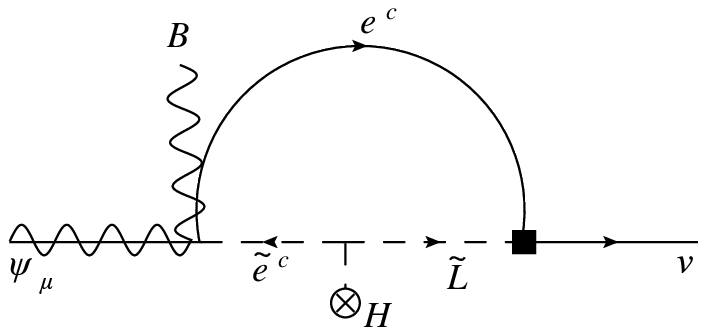}
\includegraphics[width=4.5cm, angle=0]{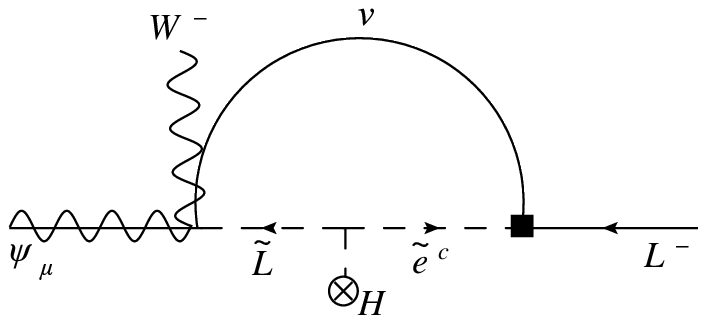}\\
\includegraphics[width=4.5cm, angle=0]{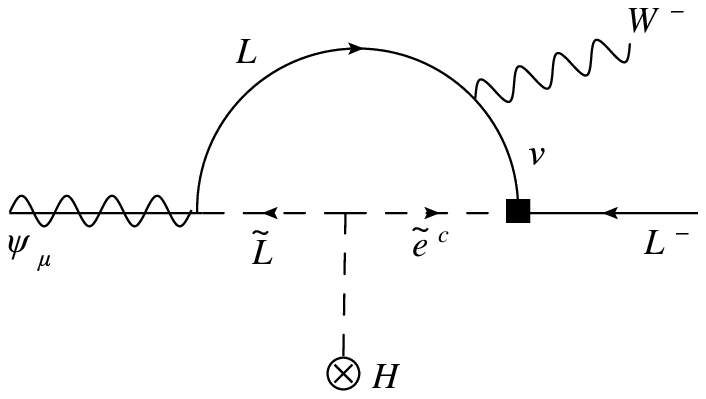}
\includegraphics[width=4.5cm, angle=0]{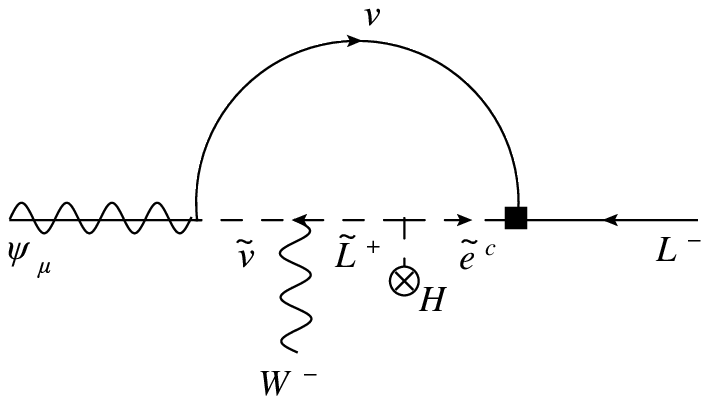}
\includegraphics[width=4.5cm, angle=0]{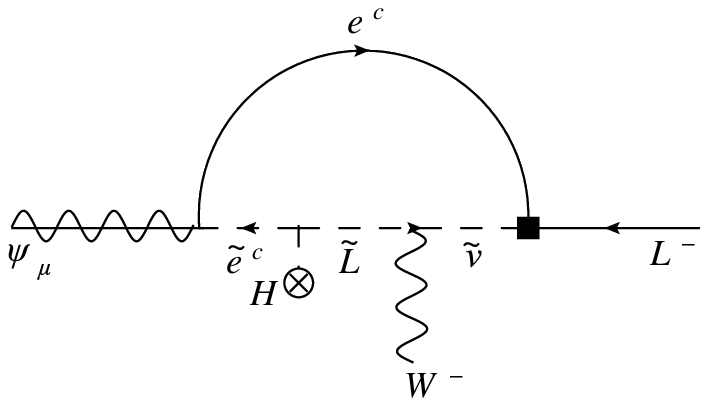}\\
\includegraphics[width=4.5cm, angle=0]{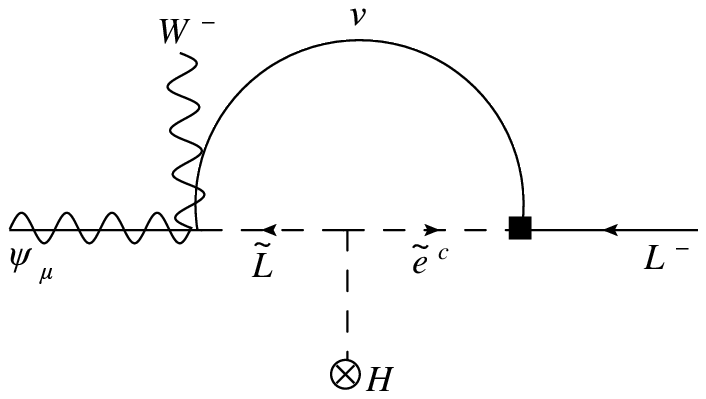}
\includegraphics[width=9cm, angle=0]{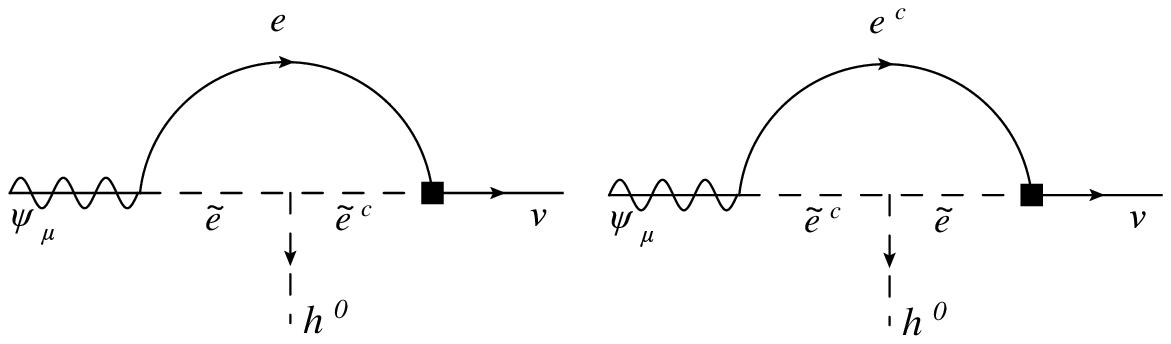}
\caption{Two-body gravitino decays through loops. Here the symbol $\otimes$ on the external Higgs fields are taken as the vacuum expectation values. 
The black boxes represent RPV couplings.}
\label{fig8}
\end{center}
\end{figure}

Second, the two-body decays are induced by the following dimension six
operators\footnote[4]{We call them dimension six because they necessarily involves the Higgs field. This is hidden in $(m_{\widetilde \ell}^2)_{LR}$ which is proportional to the Higgs VEV.}
\begin{eqnarray}\label{25}
\mathcal{L}^{\rm vertex}_{\rm eff} &=& \frac{\kappa}{6 M_{\rm Pl}}
\left[ (g/\sqrt2) i\overline{\nu} \mathbb{P}_R \gamma_\alpha
\sigma_{\mu\nu} \psi^\alpha W_3^{\mu\nu} - (g'/\sqrt2)
i\overline{\nu} \mathbb{P}_R \gamma_\alpha \sigma_{\mu\nu}
\psi^\alpha B^{\mu\nu} \right. \nonumber \\
&& \left. + g i \overline{\ell^-} \mathbb{P}_R \gamma_\alpha
\sigma_{\mu\nu} \psi^\alpha W^{-\mu\nu}\ \right] + {\rm h.c.} \ ,
\end{eqnarray}
where
\begin{eqnarray}
\kappa = \frac{\lambda (m_{\widetilde \ell}^2)_{LR}}{16\pi^2
m_{\widetilde \ell}^2} \ ,
\end{eqnarray}
which are obtained by explicit one-loop calculations of the diagrams
listed in Fig.~\ref{fig8}. In $\mathcal{L}^{\rm vertex}_{\rm eff}$, only
vertex corrections to the effective dimension six operators are
included.
There are also contributions to the gravitino two-body decays through the
gaugino-lepton mixing as shown in Fig.~\ref{fig7}. 
\begin{eqnarray}
\mathcal{L}^{\rm gaugino}_{\rm eff} &=& \frac{i}{8M_{\rm Pl}} \left[
U_{\widetilde W_3\nu} \overline{\nu} \mathbb{P}_R \gamma_\alpha
\sigma_{\mu\nu} \psi^\alpha W_3^{\mu\nu} + U_{\widetilde B\nu}
\overline{\nu} \mathbb{P}_R \gamma_\alpha \sigma_{\mu\nu}
\psi^\alpha B^{\mu\nu} + U_{\widetilde W\ell} \overline{\ell^-} \mathbb{P}_R
\gamma_\alpha \sigma_{\mu\nu} \psi^\alpha W^{-\mu\nu}\ \right] \ , \nonumber \\
\end{eqnarray}
where the lepton-gaugino mixings $U_{\widetilde
W_3\nu}$, $U_{\widetilde B\nu}$, $U_{\widetilde W\ell}$ and so are the corresponding decay amplitudes, are always
suppressed by the ratio of lepton mass to the gaugino mass, compared to the vertex
corrections in Eq.~(\ref{25}).

Third, for the sake of completeness, we also calculate the decay rate of gravitino 
to the lightest Higgs boson plus a neutrino, induced by $\lambda$. The corresponding 
Feynman diagrams are shown as the last two figures in Fig.~\ref{fig8}. After integrating out the heavy sleptons,
assuming their masses are universal (and heavy), we obtain the effective Lagrangian
\begin{eqnarray}
\mathcal{L}_{\rm eff}^{{\rm Higgs-vertex}} = \frac{m_{3/2} [ (m^2_{\widetilde \ell})_{LR}/v ] 
}{18\times(16\pi^2)M_{\rm Pl} m_{\widetilde \ell}^2}
\bar\nu \mathbb{P}_R \gamma^\mu\gamma^\nu \psi_\mu D_\nu h^0 + \rm h.c. \ ,
\end{eqnarray}
and the partial decay rate
\begin{eqnarray}
\Gamma_2(\psi_\mu\to h^0\nu) = \frac{g^2 \lambda^2}{62208\pi} \frac{
[ (m^2_{\widetilde \ell})_{LR}]^2}{(16\pi^2)^2 m_{\widetilde \ell}^4} \frac{m_{3/2}^3}{ M_{\rm Pl}^2} \frac{m_{3/2}^2}{M_W^2}
\left( 1 - \frac{m_{h^0}^2}{m_{3/2}^2} \right)^4 \ .
\end{eqnarray}

\end{document}